\documentclass[twocolumn, trackchanges]{aastex61}

\submitjournal{ApJL}

\shorttitle{Atmosphere Characterization of WASP-107\lowercase{b}}
\shortauthors{Kreidberg et al.}

\begin{document}

\title{Water, High-Altitude Condensates, and Possible Methane Depletion in the Atmosphere of the Warm Super-Neptune WASP-107\lowercase{b}}

\correspondingauthor{Laura Kreidberg}
\email{laura.kreidberg@cfa.harvard.edu}

\author{Laura Kreidberg}
\affiliation{Harvard Society of Fellows, 78 Mt. Auburn St., Cambridge, MA 02138, USA}
\affiliation{Harvard-Smithsonian Center for Astrophysics, 60 Garden St., Cambridge, MA 02138}

\author{Michael R. Line}
\affiliation{School of Earth \& Space Exploration, Arizona State University, Tempe AZ 85287, USA}

\author{Daniel Thorngren}
\affiliation{Department of Astronomy and Astrophysics, University of California, Santa Cruz, CA 95064}

\author{Caroline V. Morley}
\altaffiliation{Sagan Fellow}
\affiliation{Harvard-Smithsonian Center for Astrophysics, 60 Garden St., Cambridge, MA 02138}

\author{Kevin B. Stevenson}
\affiliation{Space Telescope Science Institute, 3700 San Martin Drive, Baltimore, MD 21218, USA}

\begin{abstract}
	The super-Neptune exoplanet WASP-107b is an exciting target for atmosphere characterization. It has an unusually large atmospheric scale height and a small, bright host star, raising the possibility of precise constraints on its current nature and formation history.  We report the first atmospheric study of WASP-107b, a Hubble Space Telescope measurement of its near-infrared transmission spectrum.  We determined the planet's composition with two techniques: atmospheric retrieval based on the transmission spectrum and interior structure modeling based on the observed mass and radius. The interior structure models set a $3\,\sigma$ upper limit on the atmospheric metallicity of $30\times$ solar. The transmission spectrum shows strong evidence for water absorption ($6.5\,\sigma$ confidence), and the retrieved water abundance is consistent with expectations for a solar abundance pattern. The inferred carbon-to-oxygen ratio is subsolar at $2.7\,\sigma$ confidence, which we attribute to possible methane depletion in the atmosphere.  The spectral features are smaller than predicted for a cloud-free composition, crossing less than one scale height. A thick condensate layer at high altitudes (0.1 - 3 mbar) is needed to match the observations. We find that physically motivated cloud models with moderate sedimentation efficiency ($f_\mathrm{sed} = 0.3$) or hazes with a particle size of 0.3 $\mu$m reproduce the observed spectral feature amplitude. Taken together, these findings serve as an illustration of the diversity and complexity of exoplanet atmospheres. The community can look forward to more such results with the high precision and wide spectral coverage afforded by future observing facilities. 
\end{abstract}


\section{Introduction} \label{sec:intro}
The composition of a planet's atmosphere depends on where and how the planet formed. By measuring the inventory of atmospheric elemental abundances, we can shed light on important aspects of the formation process such as location within the disk and the relative accretion rates of gas versus solids \citep[][]{oberg11, fortney13, madhusudhan14,  mordasini16, espinoza17}.  

The warm Neptune WASP-107b is an intriguing target for atmosphere characterization for several reasons.  It has an intermediate size between ice giants and gas giants, with a mass similar to Neptune's and a radius close to Jupiter's \citep[$0.12\,M_\mathrm{Jup}$, $0.94\,R_\mathrm{Jup}$;][]{anderson17}. Studying the atmospheres of planets in this transition region will provide additional clues in the much-debated mystery of what stunts the growth of Neptune-size planets \citep[e.g][]{pollack96, dawson16, frelikh17}.  

WASP-107b also has a relatively low equilibrium temperature compared to most other exoplanets that are amenable to atmosphere characterization (780 K, assuming zero albedo).  This results in a distinct atmospheric chemistry compared to other well-studied systems: at low temperatures, the dominant molecular reservoir for carbon transitions from carbon monoxide to methane \citep{moses13}.  Spectral features from both water and methane are accessible with current observing facilities -- both of these molecules have strong absorption bands in the wavelength range covered by \emph{HST}/WFC3.  Measuring the shape and amplitude of these spectral features enables a constraint on the abundance of water, the dominant oxygen-bearing molecule, and methane, the dominant carbon-bearing molecule, providing a bounded estimate of the carbon-to-oxygen ratio (C/O). Previous measurements of C/O have been challenging because they rely on broadband photometry or narrow wavelength coverage, and have mainly resulted in upper limits \citep[e.g.][]{madhusudhan11, line14, benneke15, kreidberg15b}. 

In addition, WASP-107b is one of the best targets discovered to date for atmosphere characterization. Thanks to its large atmospheric scale height and small, bright host star, the expected signal-to-noise for the transmission spectrum is comparable to the best studied benchmarks in the field (e.g. HD\,209458b).  In this paper we report the first atmosphere characterization of WASP-107b: a near-infrared transmission spectrum measured with the \emph{Hubble Space Telescope} (\emph{HST}; Program GO 14915, PI L. Kreidberg).

\section{Observations and Data Reduction}
We observed a single transit of WASP-107b with \emph{HST}'s Wide Field Camera 3 (WFC3) instrument on UT 5-6 June 2017.  The transit observation consisted of five \emph{HST} orbits. At the beginning of each 96-minute orbit, we took an image of the target with the F130N filter (exposure time = 4.2 s). This direct image is used for wavelength calibration. For the remainder of the target visibility period, we obtained time series spectra with the G141 grism, which provides low-resolution spectroscopy over the wavelength range $1.1 - 1.7\,\mu$m.  We used the NSAMP=6, SPARS\_25 readout mode (exposure time = 112 s) to optimize the efficiency of the observations, as determined by the \texttt{PandExo\_HST} planning tool \citep{batalha17}.  As is standard for WFC3 observations of bright targets, we used the spatial scanning observing mode, which slews the telescope in the spatial direction over the course of an exposure. The scan rate was 0.12 arcseconds/sec.

We reduced the data with the custom pipeline described in \cite{kreidberg14a}, which we summarize briefly here. For each exposure, we extracted the spectrum from each up-the-ramp sample (or ``stripe") separately using the optimal extraction algorithm of \cite{horne86}. We estimated the background from the median count level of pixels uncontaminated by the target spectrum (rows $5-250$, columns $5-515$).  The stripe spectra were then summed to create the final spectrum. For each stripe, the extraction box was 80 pixels high and centered on the stripe's midpoint in the spatial direction. We corrected the spectra for the changing dispersion solution over this aperture and small drifts over time ($<0.1$ pixel).  

\section{Light Curve Analysis}
The data analysis had two parts: the band-integrated ``white" light curve fit and the spectroscopic light curve fits.

\subsection{White Light Curve}
To create the raw white light curve, we summed each spectrum over the 181 pixels in the spectral trace.  The white light curve has systematic trends that are typical for WFC3 observations \citep{zhou17}: the flux increases asymptotically over each orbit (the ``ramp" effect) and there is a visit-long linear trend. The largest ramp occurs in the initial orbit (orbit zero), so we only fit data from orbits one through four in our analysis, following common practice.  We fit the light curve with the analytic model of the form $F_\mathrm{white}(t) = S_\mathrm{white}(t)\times T_\mathrm{white}(t)$, where $S_\mathrm{white}$ is a systematics model and $T_\mathrm{white}$ is a transit model. We used the same systematics model as \cite{kreidberg15b}.  We modelled the transit with the \texttt{batman} package \citep{kreidberg15a}.  The model parameters are the orbital period $p$, time of inferior conjunction $t_0$, transit depth $r_p/r_s$, ratio of semi-major axis to stellar radius $a/r_s$, orbital inclination $i$, and the quadratic stellar limb darkening parameters $u_1$ and $u_2$.

\begin{figure*}
\includegraphics[width = \textwidth]{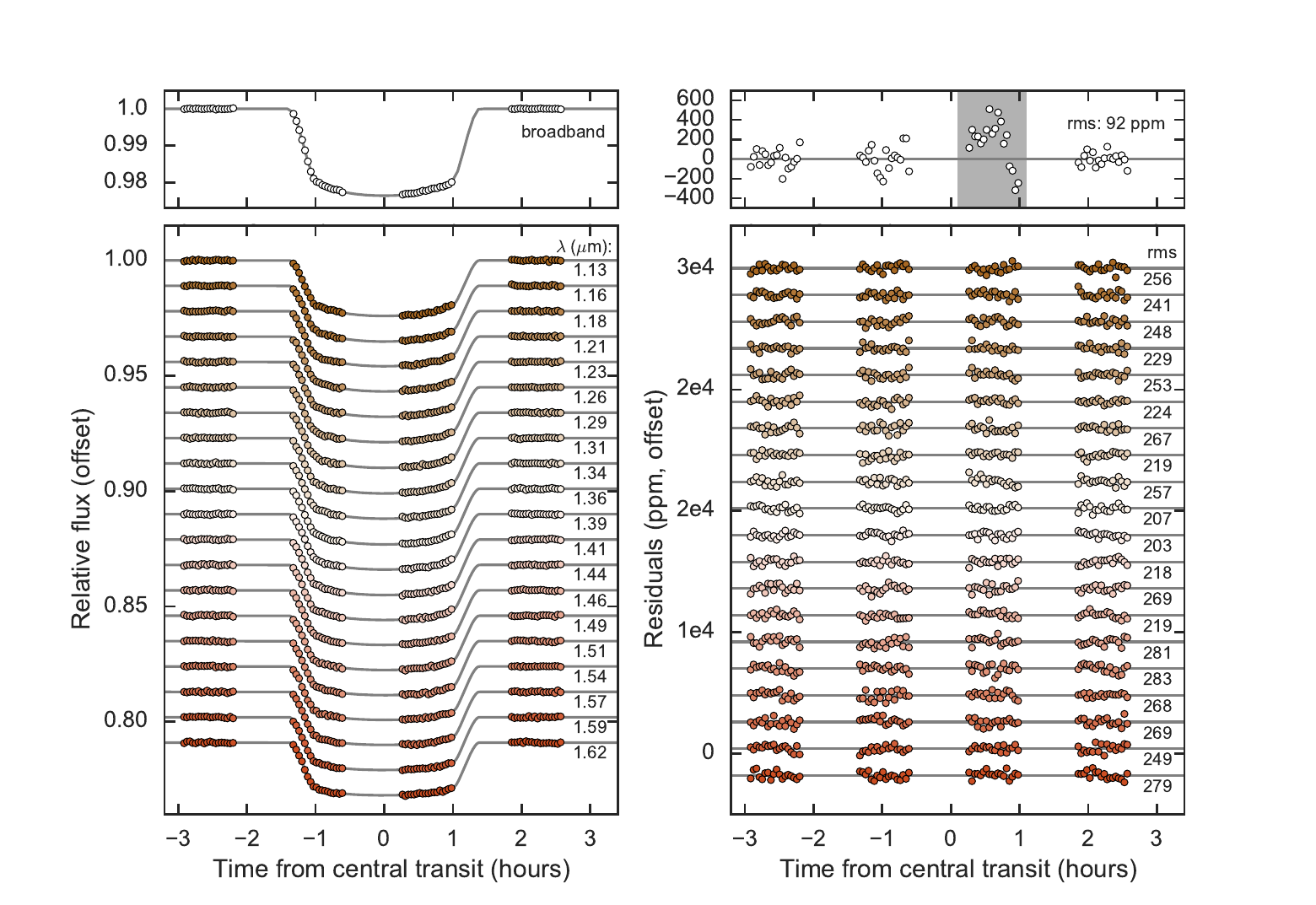}
\caption{Left: Broadband and spectrophotometric transit light curves for WASP-107b compared to best fit models. Right: residuals from the best fit. Annotations indicate the central wavelength and root-mean-square (rms) residuals. A star-spot crossing feature is shaded in gray in the upper right; our systematic error correction removes this feature from the spectroscopic light curves.} 
\label{fig:lc}
\end{figure*}

\subsubsection{Star Spot Crossing}
In our initial analysis, we noticed a bump in the light curve during orbit three that was not fit well with our model. We attribute this feature to a star spot crossing event, as WASP-107 is an active star and spot crossings have been observed before \citep{dai17, mocnik17}. In our subsequent analysis, we gave the data in orbit three no weight in the fit. The amplitude of the spot crossing feature is 300 ppm, as illustrated in Figure\,\ref{fig:lc}.

\subsubsection{Final Fit}
In our final fit, we fixed the transit parameters $a/r_s$, $i$, $p$ on the precise estimates from the Kepler light curve \citep{dai17}.  We also fixed quadratic limb darkening parameters $u_1$ and $u_2$ on predictions from a PHOENIX model for a star with effective temperature 4300 K, calculated with the \texttt{limb-darkening} package from \cite{espinoza15}.  We checked that the values we fixed were consistent with our estimates when we allowed them to vary freely.  We also checked that the uncertainty in the stellar parameters does not affect the PHOENIX model predictions at the level of precision of our data: we varied the stellar effective temperature by $\pm100$ K from the published value and found that the transmission spectrum was not significantly changed.  The remaining free parameters in the fit were $t_0$, $r_p/r_s$, and the systematics parameters for the visit-long and orbit-long trends.

For the best fit white light curve, the root-mean-square (rms) residuals were 93 ppm (excluding the star spot crossing), which is somewhat larger than the expected shot noise of 50 ppm. We attribute the excess noise to loss of flux off the edge of the detector, which can occur if there is variation in the position or length of the spatial scan. There is no evidence for correlated noise in the residuals, so to account for the excess noise we simply increased the per-point uncertainties by a factor of 1.7 to achieve a $\chi^2_\nu$ value of unity.  We then used the Markov chain Monte Carlo (MCMC) algorithm to estimate parameter uncertainties \citep{foremanmackey13}.  The chain had 50 walkers which each ran for $10^4$ steps with the first 10\% discarded as burn-in. We tested for convergence by dividing the chain in two halves and confirming that they gave consistent results. The transit time was $t_0 = 2457910.45407\pm6\mathrm{e}{-5}$ BJD$_\mathrm{TDB}$ and the planet/star radius was $r_p/r_s = 0.14399\pm0.00017$. 

\subsection{Spectroscopic Light Curve Fits}
We binned the spectrum into 20 spectrophotometric channels from 1.12 to 1.63 $\mu$m, shown in Figure\,\ref{fig:lc}. We fit the light curves with the \texttt{divide-white} technique, which assumes that the light curve systematics have the same morphology at all wavelengths \citep{stevenson14c, kreidberg14a}. For this method, the transit model $T_\lambda(t)$ is multiplied by the systematics vector from the white light curve fit ($F_\mathrm{white}/T_\mathrm{white}$), and rescaled by a factor $C_\lambda + V_\lambda t$.  One advantage of this approach is that it removes the star spot crossing feature, enabling us to use orbit three with no additional correction. The amplitude of the feature has no detectable wavelength dependence at the level of precision of our data.  As for the white light curve, we fixed some of the transit parameters on the best fit values from \cite{dai17}. We did not put put priors on the transit parameters because varying them simply shifts the spectrum up or down by a constant value. Since the planet's 10-bar radius is a free parameter in our analysis of the spectrum, the absolute transit depth does not affect the retrieved atmospheric properties. In addition to fixing the transit parameters, we also fixed the limb darkening on the PHOENIX model predictions for a star with $T_\mathrm{eff} = 4300 K$, $\log g = 4.5$, and $[\mathrm{Fe/H}] = 0.$ (coefficients listed in Table\,\ref{tab:tspec}.  The final spectroscopic light curve fits had just three free parameters per channel: $C_\lambda$, $V_\lambda$, and $r_p/r_s$.  We refer to this fitting approach as ``Method A".

\begin{figure*}
\includegraphics[width = \textwidth]{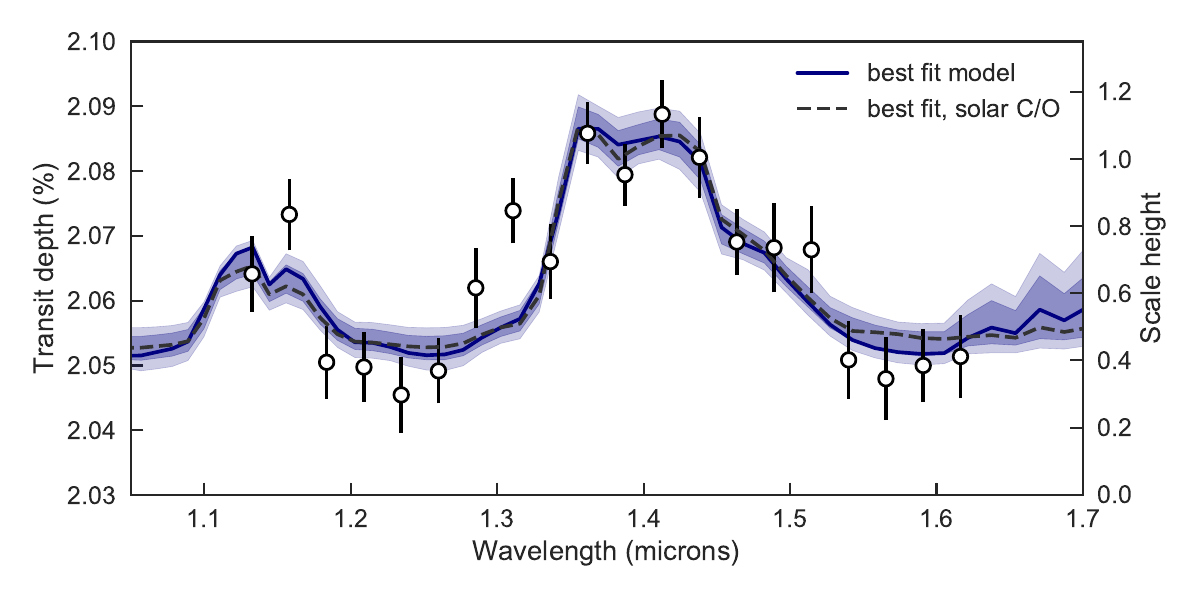}
\caption{The transmission spectrum of WASP-107b (points with 1\,$\sigma$ error bars) compared to retrieved models (blue line with shaded $1$ and $2\sigma$ confidence intervals). The features at $1.15$ and $1.4$ $\mu$m are due to water absorption. The right-hand axis indicates the normalized atmospheric scale height assuming solar composition. The water feature crosses less than one scale height, indicating that condensates are present at high altitude. We also show the best fit model from an analysis where the C/O is fixed at 0.54, the solar value (dashed line). There are subtle differences in feature shape due to methane absorption in the solar C/O case, which cause the retrieval to favor sub-solar C/O values at 2.7$\sigma$ confidence.} 
\label{fig:spectrum}
\end{figure*}

The best fit light curves have a median $\chi^2_\nu$ value of 1.16.  We
conservatively rescaled the photometric uncertainties for all spectroscopic
channels such that the $\chi^2_\nu$ values are unity. We performed an MCMC fit
to the light curves with \texttt{emcee}.  For each light curve we ran a fit with
50 walkers and 1000 steps per walker, and tested for convergence as we did for
the white light curve. The median transit depths and $1\,\sigma$ uncertainties
are given in Table\,\ref{tab:tspec}, and the transmission spectrum is shown in
Figure\,\ref{fig:spectrum}. 

We explored several alternative choices for the spectroscopic light curve fits, but found that none of them made a significant difference in the transmission spectrum. In one test, we fit the spectroscopic light curves with the same analytic model we used for the white light curve. This fit (which we label ``Method B") has additional free parameters, so the transit depth uncertainties increase by 50\%, but the best fit transit depths are consistent within $0.5\sigma$ on average with method A over the bandpass. We also tested a white light curve systematics vector $S_\mathrm{white}$ from a fit that included the star-spot crossing orbit. The results from these tests are nearly identical to our nominal transmission spectrum, except with a small constant offset due to the uncorrected star-spot crossing feature. This offset does not affect our final analysis because it is marginalized in the atmospheric retrieval with the 10-bar radius parameter (see \S\,\ref{sec:retrieval}). 

We also repeated the calculations from Method A, but fit for a linear limb darkening parameter rather than fixing the limb darkening on the PHOENIX model values. This approach, which we dub ``Method C", also yields consistent results with Method A (the relative transit depths agree to better than 0.2 sigma on average). The transit depths and fitted limb darkening coefficients are provide in Table\,\ref{tab:tspec}. The fitted limb darkening coefficients are consistent with the model predictions, so we opted to fix the coefficients in our final analysis because it improves the precision on the transit depths.  

To test that uncertainties on the stellar parameters do not significantly bias the limb darkening coefficients, we calculated PHOENIX model limb coefficients for WASP-107 over the $1\,\sigma$ range of stellar effective temperatures, surface gravities, and metallicities published in the discovery paper ($T_\mathrm{eff} = 4430\pm130$ K, $\log g = 4.5 \pm 0.1$, and $[\mathrm{Fe/H}] = 0.02 \pm 0.1$). The uncertainty in $T_\mathrm{eff}$ has the largest effect on the estimated limb darkening coefficients. We recalculated the transmission spectrum with limb darkening models that varied $T_\mathrm{eff}$ from 4310 to 4550 K. Over this temperature range, the change in limb darkening shifts the average transit depth by less than 20 ppm, which is below the level of noise in the data.  The relative change in transit depth is smaller still ($< 1$ ppm).

In addition to these tests, we also performed an independent data reduction and fit using K. Stevenson's pipeline and again found consistent results (well within $1\,\sigma$). 

\subsubsection{Impact of Unocculted Star Spots}
Unocculted star spots may influence the measured transmission spectrum \citep{mccullough14, zellem17}. To estimate the impact of unocculted spots on our data,  we first estimated the star spot properties based on the observed photometric variability in the Kepler band pass, which is 0.3\% \citep{dai17,mocnik17}. This variability can be reproduced with a single spot with covering fraction of 1\% and temperature 300 K lower than the stellar photosphere.  We calculated the effect of unocculted spots with these properties using equation (1) from \cite{mccullough14}, and predict that they will introduce a positive slope in the transmission spectrum of 8 ppm over the WFC3 bandpass, well below our measurement uncertainties. We note that cool stars may have persistent spot coverage that does not produce any photometric variability \citep{rackham18}. However, even if there is persistent spot coverage of 5\%, the spots produce a slope in the WFC3 bandpass of 50 ppm, which is still below our uncertainties.


\begin{deluxetable*}{LLLLLLL}
	\tablecolumns{7}
	\tablewidth{0pt}
	\tablecaption{WASP-107b Transmission Spectrum and Limb Darkening Coefficients\label{tab:tspec}}
	\tablehead{
		\colhead{Wavelength} & \colhead{Transit Depth} & \colhead{Transit Depth} & \colhead{Transit Depth} & \colhead{$u_1$} & \colhead{$u_2$} & \colhead{$u$} \\
	\colhead{(micron)} & \colhead{(Method A)} & \colhead{(Method B)} & \colhead{(Method C)} & \colhead{(PHOENIX)} & \colhead{(PHOENIX)} & \colhead{(fitted)}}
		\startdata
		1.121 - 1.145 & 0.020641 \pm 5.9e-05 & 0.020674 \pm 7.8e-05 & 0.020521 \pm 7.4e-05 & 0.38 & 0.13 & 0.44 \pm 0.011 \\
		1.145 - 1.171 & 0.020733 \pm 5.5e-05 & 0.020825 \pm 7.5e-05 & 0.020575 \pm 7.7e-05 & 0.37 & 0.14 & 0.41 \pm 0.012 \\
		1.171 - 1.196 & 0.020505 \pm 5.6e-05 & 0.020655 \pm 7.4e-05 & 0.020388 \pm 8.2e-05 & 0.36 & 0.14 & 0.4 \pm 0.012 \\
		1.196 - 1.222 & 0.020498 \pm 5.4e-05 & 0.020580 \pm 7.4e-05 & 0.020345 \pm 6.6e-05 & 0.36 & 0.15 & 0.41 \pm 0.012 \\
		1.222 - 1.248 & 0.020455 \pm 5.9e-05 & 0.020620 \pm 7.8e-05 & 0.020327 \pm 7.7e-05 & 0.36 & 0.15 & 0.39 \pm 0.012 \\
		1.248 - 1.272 & 0.020492 \pm 5.0e-05 & 0.020649 \pm 6.3e-05 & 0.020360 \pm 7.6e-05 & 0.35 & 0.16 & 0.39 \pm 0.01 \\
		1.272 - 1.298 & 0.020620 \pm 6.2e-05 & 0.020702 \pm 9.0e-05 & 0.020475 \pm 8.3e-05 & 0.34 & 0.17 & 0.42 \pm 0.014 \\
		1.298 - 1.323 & 0.020739 \pm 5.0e-05 & 0.020883 \pm 6.7e-05 & 0.020602 \pm 8.6e-05 & 0.34 & 0.17 & 0.39 \pm 0.011 \\
		1.323 - 1.349 & 0.020660 \pm 5.7e-05 & 0.020789 \pm 8.7e-05 & 0.020489 \pm 8.3e-05 & 0.33 & 0.18 & 0.4 \pm 0.014 \\
		1.349 - 1.374 & 0.020858 \pm 4.8e-05 & 0.020916 \pm 7.6e-05 & 0.020706 \pm 7.9e-05 & 0.32 & 0.19 & 0.42 \pm 0.012 \\
		1.374 - 1.401 & 0.020794 \pm 4.8e-05 & 0.020864 \pm 8.1e-05 & 0.020653 \pm 7.9e-05 & 0.31 & 0.20 & 0.41 \pm 0.012 \\
		1.401 - 1.425 & 0.020888 \pm 5.2e-05 & 0.020961 \pm 6.1e-05 & 0.020776 \pm 7.0e-05 & 0.30 & 0.21 & 0.41 \pm 0.01 \\
		1.425 - 1.452 & 0.020821 \pm 6.2e-05 & 0.020933 \pm 7.8e-05 & 0.020708 \pm 8.5e-05 & 0.29 & 0.21 & 0.39 \pm 0.011 \\
		1.452 - 1.476 & 0.020691 \pm 5.1e-05 & 0.020876 \pm 7.8e-05 & 0.020556 \pm 8.7e-05 & 0.28 & 0.22 & 0.36 \pm 0.012 \\
		1.476 - 1.502 & 0.020682 \pm 6.9e-05 & 0.020890 \pm 8.7e-05 & 0.020549 \pm 1.0e-04 & 0.26 & 0.23 & 0.34 \pm 0.014 \\
		1.502 - 1.528 & 0.020679 \pm 6.7e-05 & 0.020763 \pm 9.6e-05 & 0.020523 \pm 8.8e-05 & 0.25 & 0.23 & 0.37 \pm 0.015 \\
		1.528 - 1.552 & 0.020509 \pm 6.0e-05 & 0.020667 \pm 9.3e-05 & 0.020359 \pm 1.0e-04 & 0.23 & 0.25 & 0.33 \pm 0.014 \\
		1.552 - 1.579 & 0.020480 \pm 6.4e-05 & 0.020551 \pm 1.2e-04 & 0.020376 \pm 1.3e-04 & 0.22 & 0.24 & 0.38 \pm 0.019 \\
		1.579 - 1.603 & 0.020500 \pm 5.6e-05 & 0.020542 \pm 1.1e-04 & 0.020321 \pm 1.1e-04 & 0.20 & 0.24 & 0.33 \pm 0.019 \\
		1.603 - 1.629 & 0.020514 \pm 6.5e-05 & 0.020755 \pm 1.1e-04 & 0.020392 \pm 1.1e-04 & 0.19 & 0.25 & 0.29 \pm 0.019 \\
		\enddata
		\tablecomments{Our retrieval analysis uses the transit depths from Method A. Transit depths from Methods A and B use quadratic limb darkening coefficients calculated from a PHOENIX stellar model (listed in columns 5 and 6). Method C fits a linear limb darkening coefficient (fit results listed in column 7).}
	\end{deluxetable*}

\section{Composition of the Atmosphere}
In this section we discuss constraints on the composition of WASP-107b's atmosphere based on interior structure modeling and atmospheric retrieval of the measured transmission spectrum.

\subsection{Atmospheric Metallicity from Interior Structure Modeling}
\label{sec:interior}
Given WASP-107b's unusually low density, we quantitatively explored the range of atmospheric metallicities that are consistent with the observed mass and radius using the structure evolution modeling of \cite{thorngren16}.  These models assume a thermally inert heavy-element core with a convective envelope of additively mixed H/He \citep{saumon95} and heavy-element impurities.  The heavy elements were a 50-50 rock-ice mix. We evolved the planets in time using the atmospheric models of \cite{fortney07}.  The results are sensitive to assumptions about the stellar age, which is uncertain \citep[either $0.6\pm0.2$ to $8.3\pm4.3$ Gyr depending on model assumptions;][]{mocnik17}. We therefore ran two models, with uniform age priors of either $0.2-1.0$ or $1.0-13.8$ Gyr.  We used the published mass and radius estimates \citep[$0.12\pm0.01\,M_\mathrm{J}$, $0.94\pm0.02$;][]{anderson17}.  

\begin{figure}
\includegraphics[width = 0.5\textwidth]{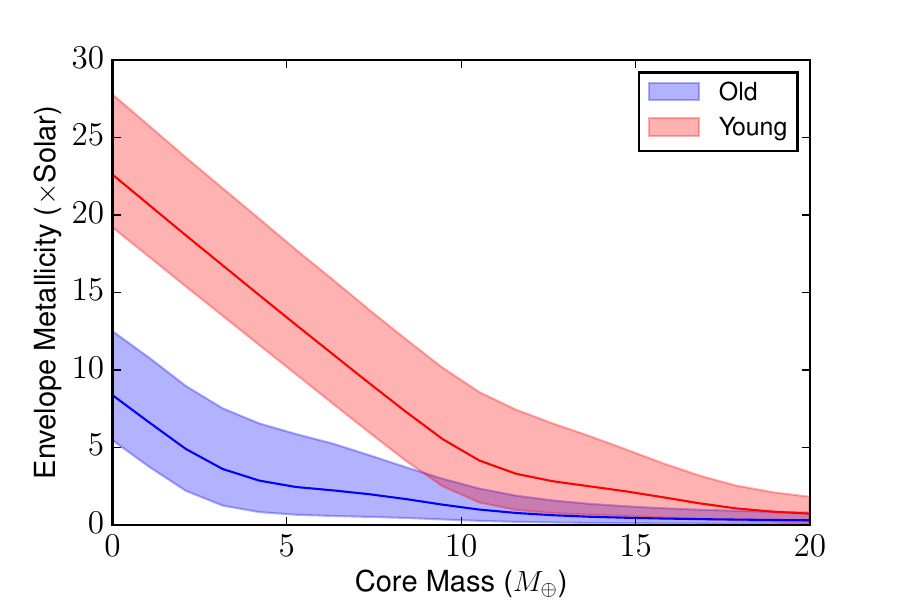}
\caption{Estimated envelope metallicity for a given core mass based on interior structure modeling. The red shading corresponds to the $1\sigma$ confidence interval for a young planet ($0.2-1$ Gyr); the blue is for an older age ($1-14$ Gyr).}
\label{fig:metal_prior}
\end{figure}

Based on these assumptions, we fit for envelope metallicity and core mass using an MCMC with uniform priors on both parameters.  The MCMC burned in for $10^3$ steps and then collected $4\times10^6$ samples.  The envelope metal mass fractions were converted to metallicities assuming the mean molecular weight of the metals was 18 (the value for water), using the approach of \cite{fortney13}. Figure\,\ref{fig:metal_prior} shows the results.  We find a $3\,\sigma$ upper limit on the metallicity of $30\times$ solar for the young stellar age range.  Higher metallicity envelopes are not allowed because they decrease the planet's radius below the observed value.  For the older age, the upper limit is even lower ($20\times$ solar), because planets cool and contract as they age \citep{fortney08}.  The largest sources of uncertainty in the metallicity estimate are the unknown core mass and stellar age, which are dominant over modeling uncertainties due to the equation of state and distribution of heavy elements in the envelope \citep{thorngren16}. By marginalizing over the unknown physical parameters, we obtain a conservative upper limit on the envelope metallicity.  Realistically, the planet probably formed with a core. Assuming a $5\,M_\oplus$ core, the upper limit on metallicity is 20 (10) $M_\oplus$ for the young (old) age.  

\subsection{Retrieval}
\label{sec:retrieval}
We also inferred the composition of the atmosphere directly from the transmission spectrum using the CHIMERA chemically-consistent retrieval \citep{line13a}.   Briefly, CHIMERA solves the transmission geometry problem using the equations in \cite{brown01, tinetti12}.  We parameterized atmospheric composition with metallicity and carbon-to-oxygen ratio under the assumption of thermochemical equilibrium using the NASA CEA routine \citep{gordon94} to compute the molecular abundances for H$_2$, He, H$_2$O, CH$_4$, CO, CO$_2$, NH$_3$, H$_2$S, Na, K, HCN, C$_2$H$_2$, TiO, VO, and FeH.    We updated the transmission model to use correlated-K opacities \citep{lacis91, molliere15, amundsen16} from the pre-tabulated line-by-line cross section database described in \cite{freedman14}. The transmission forward model is coupled with the PyMultiNest tool \citep{buchner16} to solve the parameter estimation and model selection problems.  

Our nominal model included a temperature-pressure profile (parameterized via the \citealt{guillot10} relations), the atmospheric metallicity, the C/O, a gray cloud-top pressure, and the planet's 10-bar radius.  We fixed the T-P profile morphology but scaled the irradiation temperature to allow for the unknown albedo and heat transport efficiency.  We put a uniform prior on the atmospheric metallicity of $0.01 - 30\times$ solar based on the upper limit from $\S$\,\ref{sec:interior}.  

\begin{figure}
\includegraphics[width = 0.5\textwidth]{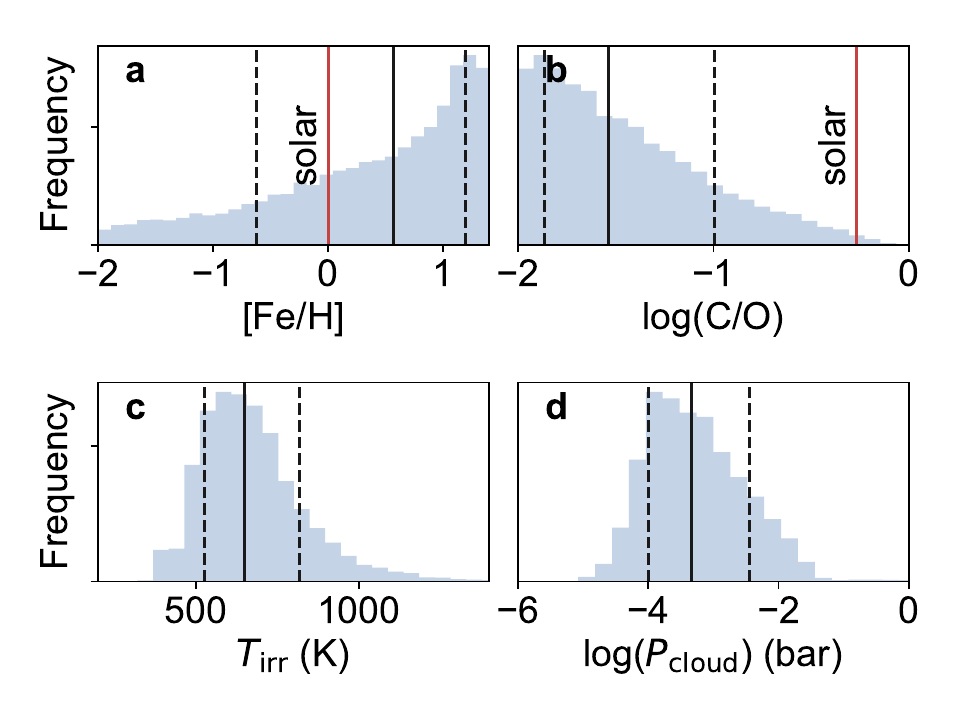}
\caption{Retrieved distributions for (a) metallicity, (b) carbon-to-oxygen ratio, (c) irradiation temperature, and (d) cloud-top pressure in bars. Black vertical lines show the median and $\pm1\,\sigma$ confidence interval (solid and dashed, respectively). Solar metallicity and C/O values are indicated with red vertical lines.}  \label{fig:retrieval}
\end{figure}

The best-fit spectrum is shown in Figure\,\ref{fig:spectrum}, and the nominal retrieval results are shown in Figure\,\ref{fig:retrieval}.  The best fit model provides a good fit to the data ($\chi^2_\nu = 1.2$).  The metallicity distribution spans the full range allowed by our priors, with preference for larger values. The cloud top pressure is estimated to be $0.01 - 3$ mbar at $1\,\sigma$ confidence. The retrieved irradiation temperature is $525 - 820$ K (1 $\sigma$ confidence).  We find that the C/O value is less than solar (0.54) at $2.7\,\sigma$ confidence.  We tested the detection significance for water by removing water opacity from the nominal model. The Bayesian evidence favors the inclusion of water at $6.5\,\sigma$ confidence. 

We explored a few retrieval scenarios with additional complexity, including the addition of cloud patchiness \citep{line16}, a quench pressure for nitrogen and carbon species \citep[e.g.][]{morley17}, a power law haze opacity, no prior on atmospheric metallicity, a more flexible T/P profile (with added free parameters for infrared opacity and visible-to-infrared opacity), and the more flexible T/P profile with quenching.  We also varied the assumed planet mass by $3\,\sigma$. These models did not significantly improve the fit quality, and the retrieved cloud-top pressure and chemical composition were consistent to within $0.1\,\sigma$.  

For example, when we removed the $30\times$ solar metallicity prior upper limit, the retrieved metallicity is $[M/H] = 1.35^{+0.68}_{-1.48}$, log10(C/O) is $-1.57^{+0.54}_{-0.31}$, and cloud top pressure $\log P_\mathrm{c} = -3.63^{+0.81}_{-0.48}.$ These results show that the inferred C/O ratio and cloud-top pressure are unaffected by the prior on atmospheric metallicity.  The metallicity is consistent with the interior structure modeling (which puts a $3\,\sigma$ upper limit on metallicity of $[\mathrm{M/H}] = 1.5$). The transmission spectrum does not constrain the atmospheric metallicity more precisely because there is a degeneracy between metallicity and the planet's 10-bar radius \citep[e.g]{griffith14,heng17} with the constraint primarily driven by the H$_2$O-H$_2$ CIA continuum ratio, molecular weight \citep{line16}, and pressure broadening \citep{dewit13}. 

Similarly, we find that the transmission spectrum does not provide a strong constraint on the shape of the temperature pressure profile. The primary constraint on temperature structure is the effective scale height temperature, which sets the amplitude of features in the spectrum. To confirm this, we explored using a more flexible T/P profile with free parameters for infrared and visible-to-infrared opacity, in addition to irradiation temperature (which we label case 2). Compared to our nominal model (case 1), which just varies irradiation temperature but keeps the T/P profile shape fixed, we found that the retrieved temperature at the photosphere is nearly identical. For example, at $P = 10^{-4}$ bar (the most probable cloud-top pressure), we find $T = 618^{+150}_{-120}$ K for case 1, and $T = 615^{+178}_{-123}$ K for case 2. The constraints on chemical composition agree at the 0.1-$\sigma$ level for the two cases.


\subsubsection{Methane Abundance}
If methane is present in the atmosphere, it has a strong spectral feature centered at $1.4\,\mu$m. This feature overlaps the water feature, but has a wider shape. Our data are precise enough to distinguish this subtle difference in feature shape, so to explore whether methane depletion is the underlying cause of the inferred low C/O, we ran two additional retrievals.  In one test, we assumed chemical equilibrium but excluded methane opacity. This set-up resulted in a much broader distribution of C/O values ($0.02 - 1.6$ at $1\,\sigma$).  We also performed a ``free" retrieval that varied the abundances of CH$_4$, H$_2$O and NH$_3$ with no assumption of chemical equilibrium. The $3\,\sigma$ upper limit on methane abundance is $1.4\times10^{-3}$. This value is in tension with expectations for a solar abundance pattern: for $3.7\times$ solar metallicity, $650$ K, and $5\times10^{-4}$ bar (the median retrieved photosphere properties), the expected methane abundance is $1.2\times10^{-3}\times$ solar (for solar C/O). Based on these tests, we conclude that the atmosphere of WASP-107b is likely depleted in methane relative to expectations for a solar abundance pattern. By contrast, the water abundance from the free retrieval ($6\times10^{-6} - 2\times10^{-3}$) is consistent with predictions for solar composition.

\subsection{Condensate Properties}
In addition to the atmospheric retrieval, we also performed forward modeling of physically motivated, self-consistent clouds and hazes using the methods described in \cite{fortney08, morley15}.  We modeled clouds that form in cool atmospheres (Na$_2$S, KCl, ZnS; see \citealt{morley12}), for a solar composition atmosphere. We varied the cloud sedimentation efficiency from 3 to 0.3 (normal to highly lofted, small particulate clouds).  Only the model with the most lofted clouds ($f_\mathrm{sed} = 0.3$) produces sufficiently small amplitude features to match the observed spectrum.  This result is in agreement with findings for other small planets, such as the low sedimentation efficiency inferred for GJ 1214b \citep[$f_\mathrm{sed} < 0.1$][]{morley15}.  


We also modeled an \emph{ad hoc} photochemical ``soot" layer near the top of the atmosphere. We predicted the abundance of hydrocarbon haze precursors from previously published $50\times$ solar photochemical models for GJ 436b \citep{line11, morley17}. We ran haze models for two different particle sizes ($0.03$ and $0.3\,\mu$m), for haze production efficiencies $f_\mathrm{haze}$ of 3, 10, and 30\%. We found that the larger particles ($0.3\,\mu$m) with moderate haze production efficiency (3\%) match the shape and amplitude of the spectral features well. More efficient haze production produces features smaller than we measure, and smaller particles sizes produce a slope (an increase in transit depth towards bluer wavelengths) that is not seen.


\section{Discussion and Conclusions} \label{sec:discuss}
We analyzed the atmospheric composition of WASP-107b based on retrieval of its near-infrared transmission spectrum and interior structure models of the planet's mass and radius.  Key results from this analysis include:

\begin{itemize}
\item{\emph{The upper limit on atmospheric metallicity from interior structure modeling is $30\times$ solar.} This limit is at the edge of consistency with the Solar System metal enrichment trend, which predicts a metallicity of $30\times$ solar for WASP-107b \citep{kreidberg14b}.  Compared to results for other exoplanets of similar mass such as GJ 436b, which has a high metallicity ($>100\times$ solar) and HAT-P-26b, which is metal-poor compared to the Solar System trend, WASP-107b adds to the evidence that exoplanets exhibit a greater diversity of compositions than is present in the Solar System \citep{morley17, wakeford17}.}
\item{\emph{The methane abundance is likely depleted relative to expectations for a solar abundance pattern, whereas water is consistent with solar composition.} This may be due to an instrinsically low carbon-to-oxygen ratio, which could arise from accretion of water-rich planetesimals \citep{mordasini16, espinoza17}.  Another possibility is that the planet has a hot interior effective temperature ($\sim500$ K), and abundances are quenched at pressure levels where CO is stable \citep[as observed in some directly imaged planets;][]{skemer14, zahnle14}. Such a high internal temperature could be due to latent heat of formation if the planet is very young, and/or tidal heating \citep{fortney08, morley17}. Further observations of the transmission spectrum over a broader wavelength range will refine the C/O estimate and help differentiate between these two scenarios.} 
\item{\emph{Optically thick condensates are present at high altitudes} ($0.1 - 3$ mbar). The amplitude of the water absorption feature in the transmission spectrum is less than a third that expected for a clear atmosphere.  We find that physically-motivated cloud and haze formation models can satisfactorily reproduce the observed feature amplitude. Either lofted clouds with low sedimentation efficiency ($f_\mathrm{sed} = 0.3$) or $0.3\,\mu$m haze particles with moderate haze production efficiency (3\%) match the spectrum. Put in context with other systems, the muted water feature for WASP-107b agrees well with the trend in feature amplitude with temperature noted in \cite{crossfield17}, indicating that condensates may be common in the atmospheres of the coolest planets.}
\end{itemize}

These results are a first look at the atmosphere of WASP-107b. The planet is already being observed at other wavelengths, including the WFC3/G102 grism and Spitzer 3.6 and 4.5 $\mu$m channels in transit and eclipse (\emph{HST} Program GO 14916, PI J. Spake, \emph{Spitzer} Program 13052, PI M. Werner; \emph{Spitzer} Program 13167, PI L. Kreidberg).  In addition, WASP-107b is included in the \emph{JWST} Guaranteed Time Observations for the NIRISS, NIRCAM, and NIRSpec instruments\footnote{\url{https://jwst-docs.stsci.edu}}.  This spate of observing programs is sure to add to the already rich and complex picture of WASP-107b's atmosphere presented here.

\acknowledgments
We thank Fei Dai, Jessica Spake, Ian Crossfield, Hannah Diamond-Lowe, and Jonathan Fortney for productive conversations. L.K. acknowledges support from the Harvard Society of Fellows and the Harvard Astronomy Department Institute for Theory and Computation. C.V.M. acknowledges support from NASA through the Sagan Fellowship Program. M.R.L. acknowledges NASA XRP grant NNX17AB56G for partial support of the theoretical interpretation of the data, as well as the ASU Research Computing staff for support with the Saguaro and Agave compute clusters.

\end{document}